# Using Decision Tree Learner to Classify Solvency Position for Thai Non-life Insurance Companies

Phaiboon Jhongpita
Technopereneurship and Innovation Management Program
Graduate School, Chulalongkorn University
Bangkok, Thailand
phaiboonj@gmail.com

Sukree Sinthupinyo
Department of Computer Engineering, Faculty of Engineering
Chulalongkorn University
Bangkok, Thailand
sukree.s@eng.chula.ac.th

Thitivadee Chaiyawat
Department of Statistics, Faculty of Commerce and Accountancy
Chulalongkorn University
Bangkok, Thailand
Thitivadee@acc.chula.ac.th

*Abstract*

*This paper introduces a Decision Tree Learner as an early warning system for classification of the non-life insurance companies according to their financial solid as strong, moderate, weak, or insolvency. In this study, we ran several experiments to show that the proposed model can achieve a good result using standard 10 fold cross-validation, split train and test data set, and separated test set. The results show that the method is effective and can accurately classify the solvency position.*

## 1. Introduction

Thai Insurance industry is subject to government regulation to safeguard the interests of the policyholders, third-party liability claimant and other related business. The number of insolvency and bankrupt Insurers can become a national issue. Solvency supervision, regulations and solvency classification are important topics for non-life insurers.

The solvency position is affected by most insurance activities and decision making process which are: premium rate making, technical reserve determining, risk undertaking, reinsurance activities, investment, sale, credibility of business to related party, country's economy, new legislations, inflation, and interest rates (Pitselis, 2009).

The office of insurance commission (OIC) of Thailand used the capital adequacy ratio (CAR) system of non-life insurance in 2009 to evaluate the capital adequacy or financial solid for the non-life insurers (See table 1). The company and regulatory actions are required if that company's total capital adequacy ratio falls below its calculation level of CAR.

In this study, two data sets are used, a 2000-2008 training data set which consists of 70 Thai non-life insurance companies and a 2009 data test set (unseen data) of 65 companies. Using ten attributes, Decision Tree Learner, resample, and SMOTE techniques are applied to classify the solvency condition.

TABLE 1. The solvency evaluation and regulatory actions based on CAR system.

| Solvency position | Capital adequacy ratio (CAR) | The action level |
|---|---|---|
| Strong | ≥ 150% | No action level |
| Moderate | 120 -150% | Company action level |
| Weak | 100 - 120% | Regulatory action level |
| Insolvency | 100% | Authorized control & Mandatory control level |

Note:
Company action level - company must file plan with insurance commissioner explaining cause of deficiency and how it will be corrected.
Regulatory action level - The commissioner is required to examine the insurer and take correct action.
Authorized control level & Mandatory control level - The commissioner has legal grounds to rehabilitate or liquidate the company, the commissioner is required to seize a company.

## 2. Literature Review

Among many empirical studies of insurance science, several studies with different techniques have been used to improve the performance of solvency and insolvency classification and prediction model. The techniques include multivariate discriminant analysis (Carson and Hoyt, 1995), logistic regression (Cummins et al., 1998), logit and probit analysis (BarNiv & Hathorn, 1997), multinomial logistic regression (Pitselis, 2009), and machine learning techniques (Brockett, et al., 1994; Salcedo-Sanz at el., 2005; Kramer, 1997, Hsiao & Whang, 2009). There is, however, a difficulty in applying these approaches in Thailand because Thai non-life insurance companies have too small number of insurer insolvencies to estimate the models.

## 3. Decision Tree Learner

In this study, we used a well-known decision tree algorithm, C4.5 (Quinlan, 1993). Decision tree algorithm is a predictive machine learning model that begins with a set of cases, and then creates a decision tree based on the attribute values of the training data that can be used to classify unseen cases.

## 4. Data and Methodology

The data set used in this study was collected from 70 non-life insurance companies in Thailand. The companies were in operation or went insolvency from 2000 to 2008. During this period, 616 cases (543 strong, 16 moderate, 13 weak and 44 insolvency) were selected as training data set as shown in Table 2. The data from year 2009 was used as a test set. The data source based on annual report of The Office of Insurance Commission (OIC) and the health insurance companies are not included on this study.

**TABLE 2** Number of Non-life Insurance used in this study

| Solvency position | 2000 | 2001 | 2002 | 2003 | 2004 | 2005 |
|---|---|---|---|---|---|---|
| Insolvency | 5 | 3 | 6 | 5 | 6 | 4 |
| Weak | 1 | 1 | 1 | 1 | 2 | 0 |
| Moderate | 0 | 1 | 1 | 2 | 1 | 4 |
| Strong | 64 | 65 | 62 | 62 | 61 | 60 |
| Total | 70 | 70 | 70 | 70 | 70 | 68 |

Note: The solvency position of non-life insurance companies from Year 2000 to 2008.

| Solvency position | 2006 | 2007 | 2008 | Total | % | 2009 |
|---|---|---|---|---|---|---|
| Insolvency | 6 | 5 | 4 | 44 | 7.1% | 6 |
| Weak | 3 | 1 | 3 | 13 | 2.1% | 1 |
| Moderate | 3 | 3 | 1 | 16 | 2.6% | 1 |
| Strong | 56 | 56 | 57 | 543 | 88.1% | 57 |
| Total | 68 | 65 | 65 | 616 | 100 % | 65 |

Note: The solvency position classified by Capital adequacy ratio (Total capital available (TCA) / Total capital required (TCR))

We collected data using 11 attributes which come from ones commonly employed in empirical studies of insurance science, and are found significant in previous studies in predicting non-life insurances' solvency (BarNiv & Hathorn, 1997; Brockett, et al., 1994; Salcedo-Sanz at el., 2005; Kramer, 1997; Hsiao & Whang, 2009; Pitselis, 2009). In this paper, we filtered attributes using the Correlation-based attribute subset evaluator and Greedy stepwise techniques. The attributes finally used in this study are presented in Table 3.

**TABLE 3** Attributes used

| | |
|---|---|
| V1 | Net premiums written / policyholders' surplus |
| V2 | Solvency margin to minimum required solvency margin |
| V3 | Policyholders' surplus & Technical reserve to net written premium |
| V4 | Claims incurred to policyholders' surplus & technical reserve |
| V5 | Gross agent's balance to Policyholders' surplus |
| V6 | Chang in policyholders' surplus |
| V7 | Investment yield |
| V8 | Investment assets to Policyholders' surplus |
| V9 | Return on total assets (ROA) |
| V10 | Loan & other investment to policyholders' surplus |
| V11 | Loss reserve & unpaid losses to policyholders' surplus |

After classifying training data set into four classes, we found that the training data set shows imbalanced class distribution, as shown in Table 2. The classification of data with imbalanced class distribution has posed a significantly low accuracy by most standard classifier learning algorithms, which assume a relatively balanced class distribution and equal misclassification costs (Sun et al., 2007).

Against the class distribution problem, we applied Resample and SMOTE (Chawla et.al., 2002) techniques in conjunction with the decision tree to reduce the imbalanced distribution problems. We employed the Resample approach to produce a random subsample of a data set using either sampling with replacement and SMOTE approach to over-sampling the minority class in this experiment.

We compared the accuracy of classification model between Resample and SMOTE approach using the data set as shown in Table 4.

**TABLE 4** Training data set after applying resample and SMOTE technique.

| Solvency position | Original data set | | Resample data set | | SMOTE data set | |
|---|---|---|---|---|---|---|
| Insolvency | 45 | 7.3% | 157 | 25.5% | 540 | 25.3% |
| Weak | 13 | 2.1% | 137 | 22.2% | 533 | 24.9% |
| Moderate | 17 | 2.8% | 144 | 23.4% | 522 | 24.4% |
| Strong | 541 | 87.8% | 178 | 28.9% | 541 | 25.3.% |
| Total | 616 | 100.% | 616 | 100.% | 2136 | 100% |

In this study, the decision tree learning was run with a confidence factor for pruning of 0.25 and a minimum number of instances per leaf of 2 using WEKA software (Hall et.al, 2009). Figure 1 shows the framework of classification process.

## 5. Experimental and Results

We used a 10 fold cross-validation and data test set (2009 data set) for the test. The results of solvency classification are shown in Table 5, 6, 7 and Table 8.

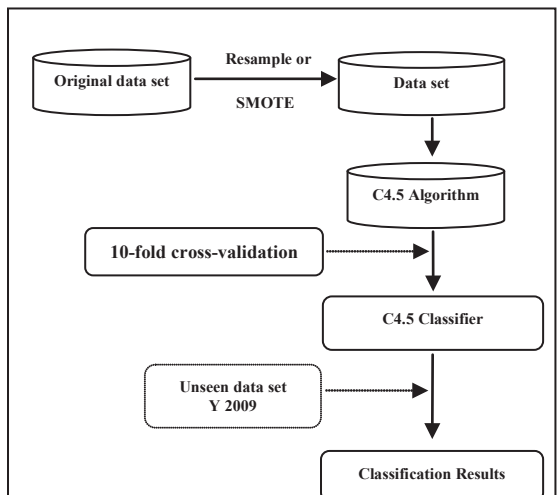

**Fig. 1** The framework of classification process

**TABLE 5** Classification results from 10 fold cross-validation (total 616 instances)- resample model

| Classification | I | W | M | S | Total | Classification Correctly (%) |
|---|---|---|---|---|---|---|
| I | 152 | 2 | 0 | 3 | 157 | 96.8% |
| W | 0 | 137 | 0 | 0 | 137 | 100% |
| M | 0 | 0 | 144 | 0 | 144 | 100% |
| S | 2 | 3 | 6 | 167 | 178 | 93.8% |
| Total | | | | | 616 | 97.4% |

I = insolvency, W = weak, M= moderate, S= strong

**TABLE 6** Classification results based on supplied test set (2009 data set) (total 65 instances)

| Classification | I | W | M | S | Total | Classification Correctly (%) |
|---|---|---|---|---|---|---|
| I | 4 | 0 | 1 | 1 | 6 | 66.7% |
| W | 0 | 1 | 0 | 0 | 1 | 100% |
| M | 0 | 0 | 1 | 0 | 1 | 100% |
| S | 1 | 2 | 1 | 53 | 57 | 93.0% |
| Total | | | | | 65 | 90.8% |

I = insolvency, W = weak, M= moderate, S= strong

**TABLE 7** Classification results based on 10-fold cross-validation (total 2136 instances) - SMOTE model

| Classification | I | W | M | S | Total | Classification Correctly (%) |
|---|---|---|---|---|---|---|
| I | 523 | 11 | 3 | 3 | 540 | 96.9 % |
| W | 5 | 513 | 15 | 0 | 533 | 96.2% |
| M | 4 | 19 | 497 | 2 | 522 | 95.2% |
| S | 4 | 0 | 28 | 509 | 541 | 94.1% |
| Total | | | | | 2136 | 95.6% |

I = insolvency, W = weak, M= moderate, S= strong

**TABLE 8** Classification results based on supplied test set (Year 2009 data set) (total 65 instances) – SMOTE model

| Classification | I | W | M | S | Total | Classification Correctly (%) |
|---|---|---|---|---|---|---|
| I | 2 | 1 | 0 | 3 | 6 | 33.3 % |
| W | 0 | 1 | 0 | 0 | 1 | 100% |
| M | 0 | 0 | 1 | 0 | 1 | 100% |
| S | 0 | 0 | 0 | 57 | 57 | 100% |
| Total | | | | | 65 | 93.8% |

I = insolvency, W = weak, M= moderate, S= strong

**TABLE 9** Performance evaluation measure

| | Resample model | | SMOTE model | |
|---|---|---|---|---|
| Cross-validation method | MAE | RMSE | MAE | RMSE |
| 10 fold cross-validation | 0.0161 | 0.1134 | 0.025 | 0.1449 |
| Supplied test set | 0.0452 | 0.2091 | 0.0326 | 0.1759 |

MAE- Mean Absolute Error
RMSE- Root Mean Squared Error

Table 9 presents a performance evaluation measure of numeric classification. In this study, we evaluated the performance of classification with MAE and RMSE method which are given by

Mean Absolute Error

$$= \frac{|p_1 - a_1| + \dots + |p_n - a_n|}{n}$$

and Root Mean Squared Error (RMSE)

$$= \sqrt{\frac{(p_1 - a_1)^2 + \dots + (p_n - a_n)^2}{n}}$$

where $P_1, P_2, \dots, P_n$ denote the predicted values on the test instances and $a_1, a_2, \dots, a_n$ represent the actual values.

From the accuracy and performance evaluation measure of numeric classification in Table 5, 6, 7, 8, and 9, the Resample method shows better performance than the SMOTE model. The Resample model was accurate classification rate of 97.4% by 10 fold cross-validation and 90.8% of the companies in a 2009 data test set. Figure 2 shows an example of the decision tree obtained from our experiment.

However, on the supplied test set (Y2009), the SMOTE model showed very good performance for weak, moderate, and strong companies (93.8%), but failed to recognize the insolvency companies.

## 6. Conclusions

We have proposed a study which applies the well-known Decision Tree Learning combined with the Resample technique which helps improve the accuracy in the case of the imbalanced distribution on the training set. The obtained results from both 10 fold cross-validation (97.4%) and supplied test set (90.8%) show the usefulness of this method and can ensure that this method can be used as an early warning system for classifying non-life insurer solvency position.

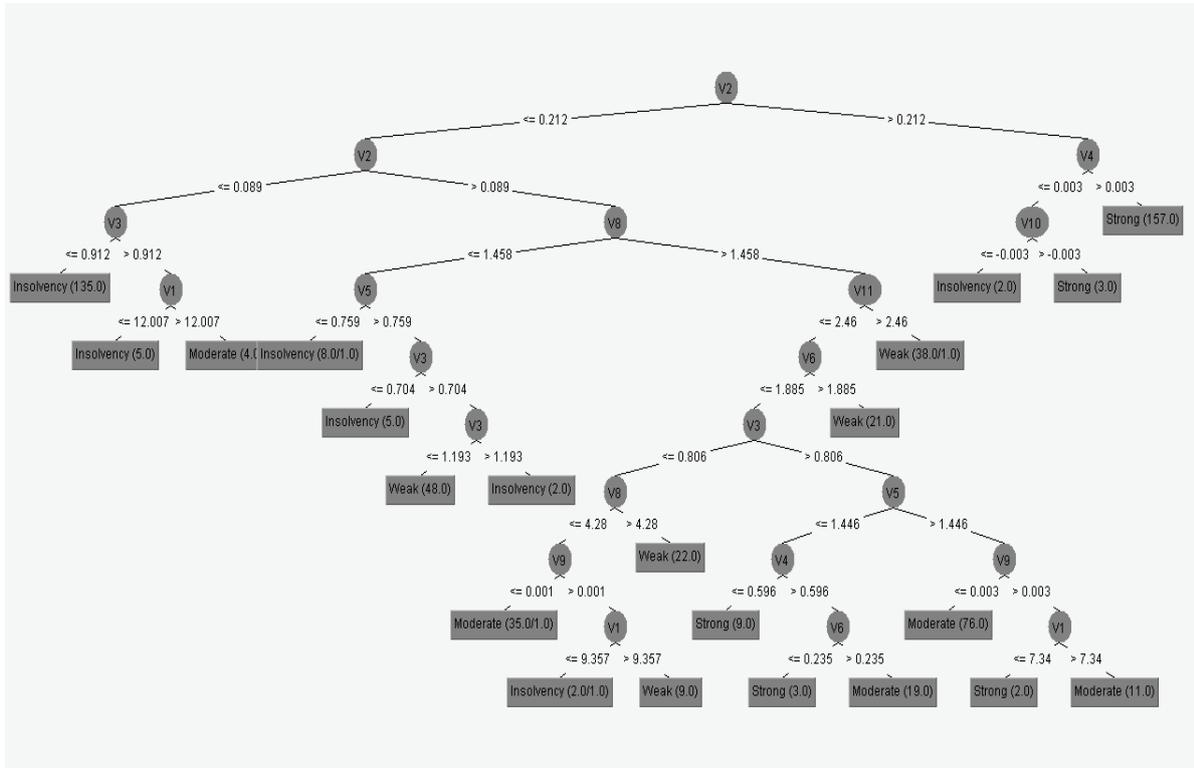

**Fig. 2** C4.5 Decision Tree: The resample model